\documentclass[a4paper,11pt]{article}
\pdfoutput=1 

\usepackage{jcappub} 

\usepackage[T1]{fontenc} 
\usepackage{graphicx}

\title{\boldmath The long freeze: an asymptotically static universe from holographic dark energy}

\author[a,1]{Samuel Blitz}
\author[b,2]{Robert J. Scherrer}
\author[b,c,3]{Oem Trivedi}

\affiliation[a]{Department of Mathematics
and Statistics, Masaryk University, Brno, Czech Republic}
\affiliation[b]{Department of Physics and Astronomy, Vanderbilt University, Nashville, TN 37235, USA}
\affiliation[c]{International Center for Space and Cosmology, Ahmedabad University, Ahmedabad 380009, India}
\emailAdd{blitz@math.muni.cz}
\emailAdd{robert.scherrer@vanderbilt.edu}
\emailAdd{oem.trivedi@vanderbilt.edu}

\abstract{We show that some holographic dark energy models
can lead to a future evolution of the universe in which
the scale factor $a$ is asymptotically constant, while
$\dot a \rightarrow 0$ and the corresponding energy and
pressure densities also vanish.  We provide specific examples of such models and general conditions that can lead to an
asymptotically static universe,
which we have called the ``long freeze."  In some cases, such evolution can follow
an arbitrarily long exponential expansion essentially
identical to the asymptotic evolution of $\Lambda$CDM.  When nonrelativistic matter is
added to the holographic dark energy, it tends to
destroy the long freeze behavior, driving the universe to recollapse.  We show that a long freeze
evolution is still possible, but only for
a more limited set of HDE models.}

\begin{document}
\maketitle
\flushbottom

\section{Introduction}
The discovery of the late-time acceleration of the Universe marked a pivotal moment in cosmology \cite{SupernovaSearchTeam:1998fmf}.  Significant efforts have been dedicated to understanding this phenomenon, exploring various methodologies. The simplest approach is a cosmological constant $\Lambda$, which,
in combination with cold dark matter (CDM), yields the
standard $\Lambda$CDM model  \cite{Weinberg:1988cp,Padmanabhan:2002ji}.
More complex solutions include modified gravity \cite{Nojiri:2010wj,Nojiri:2017ncd} and models involving scalar fields as the drivers of late-time cosmic acceleration \cite{Zlatev:1998tr,Faraoni:2000wk,Capozziello:2002rd,Odintsov:2023weg}. Furthermore, quantum gravity theories, such as braneworld cosmology in string theory, loop quantum cosmology, and asymptotically safe cosmology, have also been proposed \cite{Sahni:2002dx,Sami:2004xk,Tretyakov:2005en,Chen:2008ca,Fu:2008gh,Bonanno:2001hi}. However, persistent discrepancies remain, most notably the Hubble tension \cite{Planck:2018vyg,riess2019large,riess2021comprehensive,
Kamionkowski:2022pkx},
highlighting the limitations of our current cosmological understanding. Thus, the present cosmic epoch raises profound questions, suggesting that advances in gravitational physics could significantly enhance our current cosmological models.

Among the various solutions proposed for the late-time acceleration is the application of the holographic principle \cite{tHooft:1993dmi,Susskind:1994vu} to cosmology. This principle asserts that a system's entropy is determined by its surface area rather than its volume \cite{Bousso:1999xy}. This dark energy model has attracted interest, particularly in light of recent findings from DESI \cite{adame2024desi3,adame2024desi4,adame2024desi6}, suggesting that a deviation from $\Lambda$CDM cannot be completely dismissed. Initial research on holographic dark energy (HDE) \cite{Cohen:1998zx} indicated, through a quantum field theory (QFT) approach, that a short-distance cutoff is linked to a long-distance cutoff due to black hole formation constraints. Specifically, if $\rho$ represents the quantum zero-point energy density from a short-distance cutoff, the total energy within a region of size $L$ should not exceed the mass of a black hole of the same size, leading to the inequality $L^3\rho\leq L m_{pl}^2$, where $m_{pl}$ is the Planck
mass. Then the maximum permissible value for the infrared cutoff $L$ satisfies this inequality, yielding the relation:

\begin{equation}\label{simphde}
\rho_{HDE}=3 c^2L^{-2},
\end{equation}
where $c$ is an arbitrary parameter, and we will
take $m_{pl} = 1$ throughout. There are various choices for the cutoff scale and several extended forms of the HDE energy density beyond the simple form in Eq. \eqref{simphde}; each pair of choices corresponds to
a different HDE model.

In this paper, we show that certain HDE models can
lead to an unusual future evolution for the universe,
in which the scale factor asymptotically approaches a
constant, resulting in a static universe, which we
have dubbed the ``long freeze."
In the next section, we briefly introduce various possibilities for HDE, while in Sec. III we present a set of HDE models that result in a long freeze. In Sec. IV we discuss the implications of adding
nonrelativistic matter to the long freeze scenario, and our conclusions
are given in Sec. V.

\section{Holographic dark energy}
Numerous studies have examined holographic dark energy from various perspectives in recent years \cite{Nojiri:2017opc,Granda:2008dk,Nojiri:2005pu,Nojiri:2021iko,Nojiri:2020wmh,trivedi2024holographic,trivedi2024tsallis}. While Eq. \eqref{simphde} provides the simplest
relation between the HDE density and the cutoff $L$,
a number of other forms for $\rho_{HDE}$ as a
function of $L$ have been proposed.
For example, Tsallis HDE models incorporate Tsallis's corrections to the standard Boltzmann-Gibbs entropy, resulting in:
\begin{equation} \label{rtsa}
\rho_{HDE} = 3 c^2 L^{-(4 - 2\sigma)},
\end{equation}
where $\sigma$ is the Tsallis parameter, assumed to be positive \cite{Tavayef:2018xwx}, with the simple HDE recovered in the limit $\sigma \to 1$. Barrow's modification of the Bekenstein-Hawking formula led to Barrow HDE models described by the energy density:
\begin{equation} \label{rbar}
\rho_{HDE} = 3 c^2 L^{\Delta - 2},
\end{equation}
where $\Delta$ is the deformation parameter \cite{Saridakis:2020zol}, capped at $\Delta = 1$, and the simple HDE is regained in the limit of $\Delta \to 0$.
There are several more complex choices for the energy density as a function of $L$ that we will not discuss
here (see, e.g., Ref. \cite{Trivedi:2024dju} for
a partial listing of other proposed options).

The second component of any HDE model is the functional
form for the cutoff choice $L$.
The first HDE proposals considered a simple
Hubble horizon cutoff scale \begin{equation}\label{hcut}
    L = c H^{-1}.
\end{equation} 
Later proposals included those where the cutoff was identified with the particle horizon at time $t$,
\begin{equation}
\label{parcut}
    L_{p} = a(t) \int_{0}^{t} \frac{dt^\prime}{a(t^\prime)},
\end{equation}
or with the future cosmological event horizon \begin{equation} \label{eventcut}
    L_{f} =  a(t)\int_{t}^{\infty} \frac{dt^\prime}{a(t^\prime)}.
\end{equation}
These cutoff choices have often been riddled with problems, ranging from issues with causality to unrealistic values of the dark energy equation of state parameter.

Another choice, known as the Granda-Oliveros cutoff, was proposed as \cite{Granda:2008dk}
\begin{equation} \label{gocut}
    L = \left( \alpha H^2 + \beta \dot{H} \right)^{-1/2},
\end{equation}
where $\alpha$ and $\beta$ are constants of $\mathcal{O}(1)$.  In addition to alleviating a number of problems
with earlier proposed cutoffs, the Granda-Oliveros
cutoff had better properties with regard to
 classical stability, energy conditions, and thermodynamics.
 
 A general observation is that the results in HDE scenarios improve when given a more suitable and dynamic interplay of cosmological parameters in the cutoff scale. Another key point is that a prominent issue of the original holographic dark energy model \cite{li2004model}, in which the infrared cutoff was chosen as the size of the event horizon, is that the corresponding Friedmann equations often do not correspond to any covariant theory of gravity and may not predict the presently-observed cosmological acceleration.
 
 With these ideas in mind, Nojiri and Odintsov proposed a generalized holographic dark energy scenario.  The Nojiri-Odintsov cutoff $L$ is \cite{Nojiri:2005pu} \begin{equation} \label{gencut}
    L = L(L_{p}, \dot{L_{p}},\ddot{L_{p}},...,L_{f},\dot{L_{f}},\ddot{L_{f}},...,H,\dot{H},\ddot{H},..).
\end{equation}
It is clear from Eq. \eqref{gencut} that this cutoff scale includes, as special cases, all the previous proposals we have discussed.  It is in the context of the Nojiri-Odintsov cutoff that a long freeze can arise.

\section{The Long Freeze}
There has been recent interest in the far future evolution
of a universe dominated by holographic dark energy.
Ref. \cite{Trivedi:2024dju} showed that
big rips \cite{Caldwell:2003vq} are the most common possibility for such models, along with with pseudo rips \cite{Frampton:2011aa}, while little rip evolution \cite{Frampton:2011sp} is difficult to achieve.

Consider a far future scenario of the universe where HDE provides the total energy density.
Then the Friedmann equation is
\begin{equation} \label{friedref}
    H^2 = \frac{\rho_{total}}{3} \sim \frac{\rho_{HDE}}{3}.
\end{equation}
For the case of the Nojiri-Odintsov cutoff with
any given choice for $\rho_{HDE} (L)$
one obtains
\begin{equation} \label{basic}
    H^2 = f(L_{p}, \dot{L_{p}},\ddot{L_{p}},...,L_{f},\dot{L_{f}},\ddot{L_{f}},...,H,\dot{H},\ddot{H},..), 
\end{equation}
where the function $f$ depends on the particular choice
of model.
To keep our calculations manageable, we will
consider a simple form for the Nojiri-Odintsov cutoff
that depends only on $H$ and $\dot H$, so that
\begin{equation}
\label{Lsimple}
L = L(H, \dot H),
\end{equation}
and
\begin{equation}
\label{H2gen}
H^2 = f(H, \dot H).
\end{equation}
In some sense Eq. (\ref{Lsimple}) represents the simplest generalization
of the Granda-Oliveros cutoff.
One can then isolate $\dot H$ in the above equation
(as was done in Ref. \cite{Trivedi:2024dju}) to
obtain
\begin{equation}
\label{mainintegral}
\int_{H_i}^{H_f} \frac{dH}{g(H)} = \int_{t_i}^{t_f}dt,
\end{equation}
where $g$ is a function that can be derived from
Eq. \eqref{H2gen}.

Using this simplified form for the Nojiri-Odintsov cutoff
we now give an example of a long freeze scenario.
We will use the conventional form for $\rho_{HDE}$ as a function of $L$ (Eq. \ref{simphde}) and take the cutoff to have the form
\begin{equation} \label{vancut1}
    L = (\alpha_{1} H + \alpha_{2} H^2 + \beta \dot{H} )^{-\frac{1}{2}} 
\end{equation}
where $\alpha_{1},\alpha_{2} $ and $\beta$ are positive constants.
This cutoff has the form given in Eq. \eqref{Lsimple},
and it represents one of the simplest possible
generalizations of the Granda-Oliveros cutoff.

Substituting this expression into Eq. \eqref{simphde} gives
the energy density:
\begin{equation}
\label{vancut1den}
    \rho_{HDE} = 3 c^2 (\alpha_{1} H + \alpha_{2} H^2 + \beta \dot{H}).
\end{equation} 
From here onward we will take $c=1$, which is standard in the generalized cutoff literature. Our results
rescale in a trival way for other values of $c$.
The Friedmann equation takes the form
\begin{equation}
\label{vancut1Hubble}
    H^2 =  (\alpha_{1} H + \alpha_{2} H^2 + \beta \dot{H}),
\end{equation}
from which we derive an expression in the form of
Eq. \eqref{mainintegral}:
\begin{equation}
    \int_{H_{i}}^{H_{f}} \frac{\beta dH}{(1-\alpha_{2})H^2-\alpha_{1} H} = \int_{t_{i}}^{t_{f}} dt.
\end{equation}   
The integral gives
\begin{equation} \label{vancut1int}
    t = \frac{\beta}{\alpha_{1}} \Bigg[ \ln \left( \alpha_{2} -1 + \frac{\alpha_{1}}{H} \right)   \Bigg] +  constant,
\end{equation}
so that
\begin{equation} \label{Heq}
H = \frac{\alpha_1}{C e^{(\alpha_1/\beta)t}+ 1 - \alpha_2},
\end{equation}
where the constant $C$ is derived from the constant of integration. From this, we can get the scale factor as 
\begin{equation} \label{aa1}
    a(t)/a_{0} = \left(C+(1-\alpha_{2}) e^{-(\alpha_{1}/\beta)t}\right)^{\beta/(\alpha_{2}-1)},
\end{equation}
where $a_{0}$ is a constant. Then Eq. \eqref{vancut1den} gives
the corresponding energy density,
\begin{equation} \label{r1}
    \rho_{HDE} (t) = \frac{3 \alpha_{1}^2}{\left(1-\alpha_{2}+C e^{(\alpha_{1}/\beta) t}\right)^2}.
\end{equation}
The pressure $p_
{HDE}$ is related to the density via
\begin{equation}
\label{rhoevol}
\dot \rho_{HDE} + 3H(\rho_{HDE} + p_{HDE}) = 0,
\end{equation}
from which we have
\begin{equation} \label{p1}
    p_{HDE}(t) = \frac{(2 \alpha_{1}^2/\beta) C e^{(\alpha_{1}/\beta) t}-3 \alpha_1^2}{  \left(1-\alpha_{2}+C e^{(\alpha_{1}/\beta) t}\right)^2}.
\end{equation}
Finally, the equation of state parameter becomes \begin{equation} \label{w1}
    w = \frac{p}{\rho} = \frac{2 C e^{(\alpha_{1}/\beta) t}}{3 \beta }-1.
\end{equation}
Note that the value of the constant $C$ can be determined from Eq. (\ref{Heq}) using the value of $H$ at a fiducial initial value of $t$.

The asymptotic evolution in this case is quite
interesting.  As $t \rightarrow \infty$, the Hubble parameter, the energy density, and the pressure all vanish,  while the scale factor goes to a constant.
Such a long freeze scenario, in which the scale factor approaches a constant as $t \rightarrow \infty$, has been discussed previously in the context of other models.
Kouwn et al. \cite{Kouwn:2011qm} showed that such
behavior can arise in a standard, albeit complex Friedmann model with a
canonical scalar field, a phantom scalar field, cold
dark matter, and a negative cosmological constant, while a similar model was developed in Ref. \cite{El-Nabulsi:2015mba}.  Liu and Piao \cite{Liu:2012hr} proposed an asymptotically static universe by constructing
specific forms for $H(t)$ and then deriving a scalar
field model corresponding to one such form, although their
models require the scalar field component to have negative
energy density. Furthermore, in Ref. \cite{Guberina:2002wt} renormalization group runnings of the cosmological constant were considered to create an asymptotically constant scale factor scenario in very particular parameter ranges.  

In general, however, it is extremely
difficult to derive cosmological models that asymptote to
a constant scale factor in the context of the standard
Friedmann equation \cite{Scherrer:2022nnz}.  For example,
the loitering universe \cite{Sahni:1991ks}, which
contains a positive cosmological constant and positive
curvature, can be fine-tuned to allow $a$ to be nearly constant for an arbitrarily long time; however, it inevitably transitions into a final phase of exponential
expansion driven by the cosmological constant.  Our results suggest that future long freeze evolution can occur more naturally in the context of HDE.

One unusual result is the asymptotic behavior of the equation of state parameter, $w$.  From Eq. (\ref{w1}), we see that in the long-time limit,
we have $w \rightarrow \infty$.  This might seem
impossible, since the first and second derivatives
of the scale factor are
are given by
\begin{eqnarray}
\label {a1}
(\dot a/{a})^2 &=& \rho/3, \\
\label {a2}
\ddot a/{a} &=& -\frac{1}{6}\rho (1+3w). 
\end{eqnarray}
In the standard loitering universe scenario, one achieves
$\dot a = \ddot a = 0$ by combining a cosmological constant
($w = -1$), nonrelativistic matter ($w = 0$) and curvature
($w = -1/3$) so that the right-hand sides of both Eqs. (\ref{aa1}) and ($\ref{a2}$) are zero \cite{Sahni:1991ks}.  Then in Eq. (\ref{a2}),
the total effective equation of state parameter is
$w = -1/3$.  

How, then, can we have an asymptotically
static universe with $w \rightarrow \infty$?  The answer
is that $w = -1/3$ is a sufficient but not a necessary
condition to obtain $\ddot a = 0$.  From Eq. (\ref{a2}),
an alternative route to this result is for $\rho \rightarrow 0$.  It is instructive to rewrite
the behavior of $\rho_{HDE}$ and $p_{HDE}$ as
functions of the scale factor.  Then
in the example given above, in the limit where
$t \rightarrow \infty$, we have the asymptotic behavior
\begin{eqnarray}
\rho_{HDE} \sim \ln^2(a_f/a),\\
p_{HDE} \sim \ln (a_f/a),\\
H \sim \ln(a_f/a),
\end{eqnarray}
where $a_f$ is the asymptotic value of $a$ in the long-time limit.  It is easy to see that
these expressions
yield a long freeze and satisfy equation (\ref{rhoevol}), yet they correspond to $w \rightarrow \infty$, rather than $w \rightarrow -1/3$. In principle, a fluid satisfying these equations would yield a long freeze for the standard Friedmann expansion.  The fact that this possibility has not been previously explored in the literature can be ascribed to the fact that these equations correspond to a very unnatural equation of state for a barotropic fluid.  As we have shown, this result arises much more naturally in the context of holographic dark energy.

Now consider the general conditions on HDE models needed to produce a long freeze.  First note that $H \rightarrow 0$ as $t \rightarrow \infty$ is a necessary but not sufficient condition for $a$ to evolve to a constant.  For instance, a matter or radiation dominated universe has, respectively, $a \propto t^{1/2}$ or $a \propto t^{2/3}$, corresponding to $H = 1/2t$ or $H = 2/3t$.  Clearly, in the long time limit, $H \rightarrow 0$ and $\rho \rightarrow 0$, but $a$ increases forever.  Instead, we require that
\begin{equation}
\label{Hcond}
\int H dt \rightarrow constant,
\end{equation}
as $t \rightarrow \infty$, since this integral is just equivalent to $\ln a$.  Then from Eq. (\ref{Hcond}), as
$t \rightarrow \infty$, $H$ must go to zero more rapidly than
$1/t$.

Now consider a cutoff $L$ that generalizes Eq. (\ref{vancut1}), namely
\begin{equation}
\label{Lgeneral}
L = (\beta \dot H + f(H))^{-1/2}
\end{equation}
where $f(H)$ is an arbitrary function of $H$.  Clearly this
is not the most general possible form for $L$ consistent with
Eq. (\ref{Lsimple}), but it includes a wide range of possibilities
and will provide insight into the conditions needed for a long
freeze.  Combining this expression for $L$ with the simplest
expression for $\rho_{HDE}$ (Eq. \ref{simphde}) and again setting $c=1$
gives
\begin{equation}
\rho_{HDE} = 3(\beta \dot H + f(H)),
\end{equation}
so that
\begin{equation}
\label{genmodel}
H^2 = (\beta \dot H + f(H)).
\end{equation}
Then we have
\begin{equation}
\label{keyintegral}
    \int_{H_{i}}^{H_{f}} \frac{\beta dH}{H^2- f(H)} = \int_{t_{i}}^{t_{f}} dt.
\end{equation}   
The existence of a long freeze as $t \rightarrow \infty$
is determined by the behavior of the denominator on the left-hand side in the limit when $H \rightarrow 0$.

\begin{figure}[!h]
    \centering
    \includegraphics[width=1\linewidth]{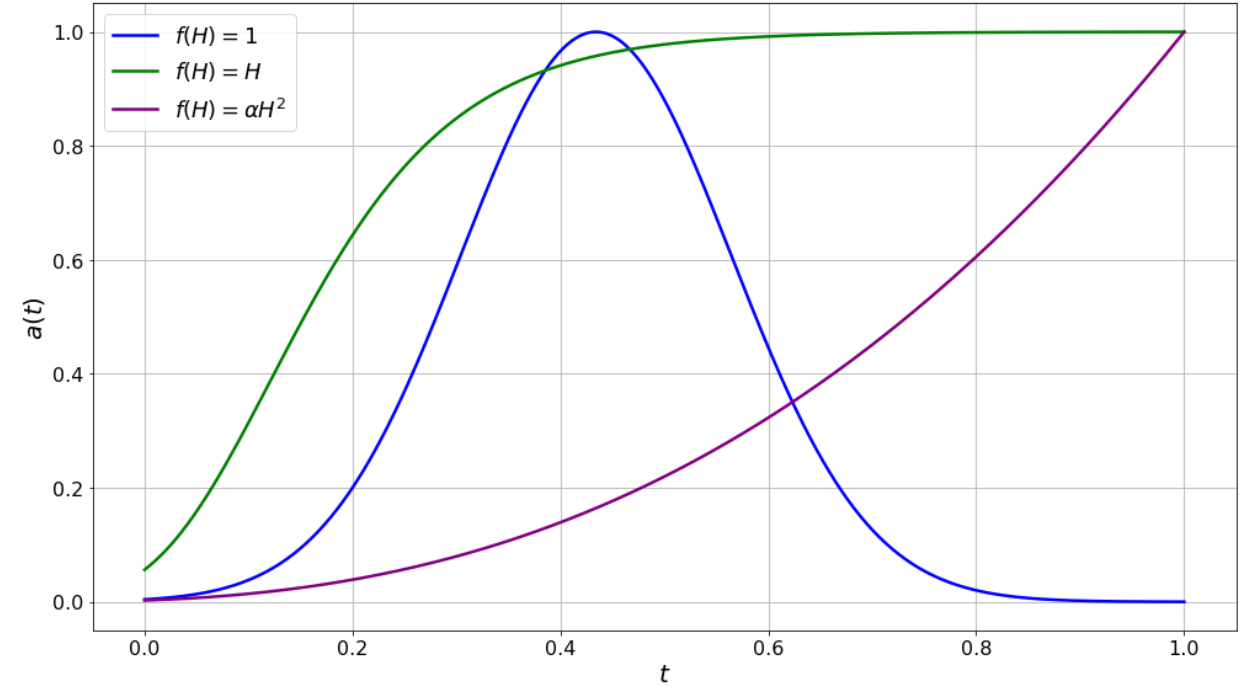}
    \caption{The scale factor $a$ as a function of the time $t$, both in arbitrary units, for the HDE model given by Eq. (\ref{genmodel}) with the indicated
    functional forms for $f(H)$.}
    \label{combplot}
\end{figure}

We will assume
that $f(H)$ scales as some power of $H$ as $H \rightarrow 0$, namely $H \sim H^n$.  A long freeze requires $\dot H < 0$,
so that $H^2 < f(H)$ as $H \rightarrow 0$.  This is clearly
impossible for $n > 2$.  For $1 < n < 2$, we obtain, in the limit where $H \rightarrow 0$, the asymptotic behavior
$H \sim t^{1/(1-n)}$, corresponding to
$a \sim \exp(t^{(2-n)/(1-n)})$.  Thus, in this case,
$H \rightarrow 0$ and $ a \rightarrow constant$ as
$t \rightarrow \infty$, corresponding to a long freeze.
For $n < 1$, $H$ evolves to negative values, indicating
a universe that expands to a maximum value of $a$ and recollapses.  The two special cases $n=1$ and $n=2$ correspond
to two cases discussed above.  For $n=1$, we have the specific
long freeze model derived from the cutoff in Eq. (\ref{vancut1}).  The $n=2$ case can lead
to several different trajectories depending
on whether $\dot H$ is positive, negative, or zero in Eq. (\ref{genmodel}).  The case
$\dot H < 0$ with $n=2$
gives the power law evolution
discussed previously and does not correspond to a long freeze, while $\dot H \ge 0$ produces
accelerated expansion.  This behavior is illustrated
in Fig. \ref{combplot} for the cases $n=0,1,2$,
where we have taken $\dot H > 0 $ for $n=2$.

Thus, our final condition for a long freeze is
that, as $H \rightarrow 0$, the function $f(H)$ 
in Eq. (\ref{Lgeneral}) scales as $H^n$, with $1 \le n < 2$.  While this condition may
seem rather restrictive, it actually corresponds to a wide
variety of functions.  Any Taylor expansion of $f(H)$ around
$H=0$ that does not contain a constant and does contain a linear term will satisfy our condition for a long freeze.

\section{Adding Nonrelativistic Matter}

Our discussion thus far has assumed a universe
containing only HDE, with no other components.
However, even if the universe undergoes a long period
of future exponential expansion, we would expect
a small relic abundance of nonrelativistic matter
to remain.  In the section, we examine
the evolution of our long-freeze HDE models when nonrelativistic matter contributes
to the expansion as well.
We will show that
for a broad class of dark energy models, the presence of even an infinitesimal matter density inevitably leads to a finite-time cosmological recollapse. At the same time, we identify a class of models where the long freeze can coexist with late-time matter contributions.

Consider a general holographic dark energy
density given by
\begin{equation}
    \rho = 3f(H, \dot{H}),
\end{equation}
where we take $f$ to be continuous,
so that divergences in $\rho$ can only arise through divergences in $H$ or $\dot{H}$. Furthermore, we restrict attention to functions $f$ that produce a long freeze in the absence of matter for all choices of expanding initial conditions, excluding finely tuned cases that require specific initial data.

Writing $s = \log a$, the Friedmann equation becomes
\begin{equation}
    \dot{s}^2 = f(\dot{s}, \ddot{s}),
\end{equation}
with a long freeze requiring that $\dot{s} \rightarrow 0$ and $\ddot{s} \rightarrow 0$ as $t \rightarrow \infty$, leading to
\begin{equation}
    \lim_{t \rightarrow \infty} f(\dot{s}, \ddot{s}) = 0,
\end{equation}
which, by continuity of $f$, implies
\begin{equation}
    f(0, 0) = 0.
\end{equation}

Now consider what happens when we add nonrelativistic
matter.
If a nonzero matter density is added, the Friedmann equation is modified to
\begin{equation}
    \dot{s}^2 = f(\dot{s}, \ddot{s}) + \dot{s}_0^2 \Omega_{m0} e^{-3(s-s_0)},
\end{equation}
where $\Omega_{m0}$ is the fraction of the total
energy density (assumed to consist entirely of
HDE and nonrelativistic matter) in the form of
nonrelativistic matter at the fiducial value of
$s = s_0$.
However, this equation is inconsistent in the long freeze limit where $\dot{s} \rightarrow 0$ and $s \rightarrow \mathrm{const.}$, as
\begin{equation}
    \lim_{t \rightarrow \infty} \dot{s}_0^2 \Omega_{m0} e^{-3(s-s_0)} \neq 0.
\end{equation}
Hence, the long freeze is destabilized by the presence of matter.

We now demonstrate that this instability results in a big crunch. The possibilities for the late-time evolution are unbounded growth of the scale
factor, oscillation, or recollapse. Growth is excluded, as in the late-time limit the Friedman equation gets arbitrarily close to the matter-free scenario which, as prescribed, should instead asymptote to a long freeze---a contradiction. This implies that the universe must eventually turn around and begin to contract.
Oscillatory solutions are inconsistent with uniqueness of physical solutions to the Friedman equation. Specifically, oscillatory solutions require the existence of some time $t_0$ such that $\dot{s}(t_0) = 0$ and $\ddot{s}(t_0) > 0$. Treating this time as the initial condition for an initial value problem, the Friedman equation reduces to the matter-free scenario, which by uniqueness must have a constant solution---another contradiction. Therefore, collapse is inevitable.

To establish that this recollapse occurs in finite time, we suppose for contradiction that the scale factor is defined for all time. If $\dot{s}^2$ grows slower than $e^{-3s}$, then $f(\dot{s}, \ddot{s})$ must cancel the matter term asymptotically, which introduces fine-tuning. If $\dot{s}^2$ grows as $A e^{-3s}$ for some constant $A >0$, then we can solve the the Friedman equation asymptotically, yielding
\begin{equation}
    s(t) = \tfrac{2}{3} \ln \left( e^{3 s(t_0)/2} - \tfrac{3}{2} \sqrt{A} (t-t_0) \right)\,,
\end{equation}
which diverges at finite $t > t_0$---contradicting existence for all time. Finally, if $\dot{s}^2$ grows faster than $e^{-3s}$, then $s \rightarrow -\infty$ even in the absence of matter, contradicting the assumption that $f$ leads to a long freeze.
Hence, in any model where $f$ is continuous and supports a long freeze in the absence of matter, any nonzero $\Omega_{m0}$ drives the universe to a big crunch in finite time.

Note that a long freeze is not impossible
to achieve in the presense of nonrelativistic matter, but it requires us to give up the assumption
that $f(H, \dot H)$ is a continuous function.
Consider, for example, the functional form
\begin{equation}
    f = H^2 + \frac{\alpha H}{\dot{H}},
\end{equation}
which is discontinuous at $\dot{H} = 0$. The corresponding Friedmann equation with matter is
\begin{equation}
    \frac{\alpha H}{\dot{H}} = - {H_0^2 \Omega_{m0}}(a/a_0)^{-3},
\end{equation}
which can be solved exactly:
\begin{equation}
    a(t)/a_0 = \sqrt[3]{\frac{ \alpha + 3 H_0^3 \Omega_{m0}}{\alpha e^{\frac{\alpha + 3 H_0^3 \Omega_{m0}}{H_0^2 \Omega_{m0}} t} + 3 H_0^2 \Omega_{m0}}} \;e^{\frac{\alpha + 3 H_0^3 \Omega_{m0}}{3 H_0^2 \Omega_{m0}} t}.
\end{equation}
This solution asymptotes to a constant:
\begin{equation}
    \lim_{t \rightarrow \infty} a(t)/a_0 = \left(\frac{\alpha + 3 H_0^3 \Omega_{m0}}{\alpha} \right)^{1/3},
\end{equation}
implying a stable long freeze. Larger $\Omega_{m0}$ leads to a larger final scale factor because
\begin{equation}
    \dot{H} = -\frac{\alpha H}{\Omega_{m0}(t)},
\end{equation}
so higher matter density slows the approach to $H = 0$.
This model generalizes to HDE densities of the form
\begin{equation}
    f(H,\dot H) = \alpha H + \beta H^2 + \gamma \frac{H}{\dot{H}},
\end{equation}
which can lead to a long freeze even in the presence of matter.
Since $f$ is not continuous, our earlier arguments excluding long freeze stability with matter do not apply.

\section{Conclusions}
We have demonstrated that, under very general assumptions,
the HDE model can lead naturally to a long freeze, in which the scale factor asymptotically approaches a constant.  In particular, this behavior can arise in the context of the generalized Nojiri-Odintsov cutoff.  However, it is also clear from the discussion in Sec. III that it cannot occur for the Granda-Oliveros cutoff or the simple Hubble horizon cutoff.  

It is interesting to note that in some of these long freeze models, the long freeze can be preceded by a period of exponential expansion.  Consider, for example, the specific cutoff given in Eq. (\ref{vancut1}) for the case $\alpha_2 = 1$.
In this case, Eq. (\ref{Heq}) gives
\begin{equation}
\label{Hexp}
H = H_0 \exp(-\alpha_1 t/\beta),
\end{equation}
where $H_0$ is the value of $H$ at the
initial time $t=0$.  Then the scale factor is given
by
\begin{equation}
\label{aexp}
a = \exp\left[(\beta/\alpha_1)H_0( 1- \exp(-\alpha_1 t/\beta)\right],
\end{equation}
where we are taking $a = 1$ at $t=0$.

Consider the evolution of the universe given by Eqs. (\ref{Hexp}) and (\ref{aexp}).  At early times, for $t \ll \beta/\alpha_1$, we have $H \approx H_0$, and $a \approx \exp(H_0 t)$, corresponding to exponential expansion.  Then, when $t \gg \beta/\alpha_1$, the scale factor asymptotically approaches the constant value $a = \exp(\beta H_0/\alpha_1)$.  Hence, we have here exponential
expansion followed by a long freeze.

It is important to note that a long freeze tends to be destabilized by the presence of nonrelativistic matter.  In a wide class of models, the addition of matter converts the long freeze scenario into a future recollapse instead.
While it is possible to construct models (as in the previous section) which contain both HDE and matter and which evolve to a future long freeze, such
models are based on somewhat more contrived forms
for the cutoff.

In this paper we have considered only the simplest form for the density as a function of the cutoff scale (Eq. \ref{simphde}).  It is straightforward to extend our arguments to more complicated forms for $\rho_{HDE}(L)$, such as the Tsallis and Barrow HDE models, and it is possible
that such models might lead to a long freeze in the presense of matter with less contrived forms for the
cutoff scale.

The long freeze provides a unique cosmic scenario. Unlike finite-time singularities (such as the big rip or big brake) that feature blow-ups in cosmological parameters, the long freeze represents an asymptotic approach to a static state where the expansion rate freezes. This distinction makes the long freeze a unique and viable alternative to the standard late-time cosmology, providing a physically meaningful scenario that warrants further exploration.
	
	\acknowledgments

The authors would like to thank Sergei Odintsov and Sunny Vagnozzi for helpful discussions.


\bibliography{JSPJMJcitations.bib}

\bibliographystyle{unsrt}

\end{document}